\def \AAP #1 #2 {{\em Astron. Astrophys.\/} {\bf #1}, #2}
\def \AAL #1 #2 {{\em Astron. Astrophys. Lett.\/} {\bf #1}, L#2}
\def \AAR #1 #2 {{\em Astron. Astrophys. Rev.\/} {\bf #1}, #2}
\def \AAS #1 #2 {{\em Astron. Astrophys. Suppl. Ser.\/} {\bf #1}, #2}
\def \AJ #1 #2 {{\em Astron. J.\/} {\bf #1}, #2}
\def \ANNREV #1 #2 {{\em Ann. Rev. Astron. Astrophys.\/} {\bf #1}, #2}
\def \APJ #1 #2 {{\em Astrophys. J.\/} {\bf #1}, #2}
\def \APJL #1 #2 {{\em Astrophys. J. Lett.\/} {\bf #1}, L#2}
\def \APJS #1 #2 {{\em Astrophys. J. Suppl.\/} {\bf #1}, #2}
\def \APSS #1 #2 {{\em Astrophys. Space Sci.\/} {\bf #1}, #2}
\def \ASR #1 #2 {{\em Adv. Space Res.\/} {\bf #1}, #2}
\def \BAIC #1 #2 {{\em Bull. Astron. Inst. Czechosl.\/} {\bf #1}, #2}
\def \JSQRT #1 #2 {{\em J. Quant. Spectrosc. Radiat. Transfer\/} {\bf #1}, #2}
\def \MN #1 #2 {{\em Mon. Not. R. Astr. Soc.\/} {\bf #1}, #2}
\def \MEM #1 #2 {{\em Mem. R. Astr. Soc.\/} {\bf #1}, #2}
\def \PLR #1 #2 {{\em Phys. Lett. Rev.\/} {\bf #1}, #2}
\def \PASJ #1 #2 {{\em Publ. Astron. Soc. Japan\/} {\bf #1}, #2}
\def \PASP #1 #2 {{\em Publ. Astr. Soc. Pacific\/} {\bf #1}, #2}
\def \NAT #1 #2 {{\em Nature\/} {\bf #1}, #2}
\def \SAIT #1 #2 {{\em Mem.\ Soc.\ Astron.\ It.\/} {\bf #1}, #2}
\def \MESS #1 #2 {{\em The Messenger\/} {\bf #1}, #2}
\def \ASTRNACH #1 #2 {{\em Astron. Nach.\/} {\bf #1}, #2}

\def\kms{{\rm\,km\,s^{-1}}}

\def\msun{M_\odot}  

\def\be{\begin{equation}}
\def\ee{\end{equation}}
\def\baray{\begin{eqnarray}}
\def\earay{\end{eqnarray}}

\def\mmc{{\rm mmc}}

\def\simless{\mathbin{\lower 3pt\hbox 
   {$\rlap{\raise 5pt\hbox{$\char'074$}}\mathchar"7218$}}} 
\def\simgreat{\mathbin{\lower 3pt\hbox
   {$\rlap{\raise 5pt\hbox{$\char'076$}}\mathchar"7218$}}} 

\documentstyle{BSAXPwork}
\input epsf.sty
\begin{opening}
\title{Spin-perpendicular kicks from evanescent binaries
formed in the aftermath of rotational  core-collapse and  
the nature of the observed bimodal   distribution of 
pulsar peculiar velocities}
\author{M. Colpi$^{1}$, \& I. Wasserman$^2$}
\institute{$^1$Department of Physics, University of Milano Bicocca, Milan,
Italy\\
$^2$Center for Radiophysics and Space Research, Cornell University, Ithaca, USA}
\date{} 
\end{opening}

\begin{document}

\oddpagefooter{}{}{} 
\evenpagefooter{}{}{} 
\medskip  

\begin{abstract} 
We argue that if core collapse leads to the formation of a rapidly
rotating proto-neutron star core or a fizzler
surrounded by fall-back material, a lighter proto-neutron star 
forms around the main star moving in a super-close orbit, as an end result
of a fully developed dynamical non-axisymmetric instability.  
Tidal mass exchange (even through a common
envelope phase)  propels
the lighter star toward the minimum stable mass for a 
proto-neutron star, whereupon it explodes and the short-lived binary disrupts.
The star that remains,
a newly born neutron star (or a black hole) acquires  
a {\it recoil velocity}  $V_{\rm kick}$, 
according to the law of conservation of linear
momentum.  
A noteworthy feature of this process is that the final kick is 
determined by nuclear physics, 
and produces in reality a widespread range in peculiar velocities, up to the 
{\it highest values} observed in the pulsar sample $\simgreat 1600 \kms$.  
Interestingly, $V_{\rm kick}$ scales with the mass
$M$ of the star that remains as $M^{-2/3}.$ The {\it kick,} lying in 
the orbital plane of the binary, is expected to be nearly
{\it perpendicular 
to the spin vector}  of the post-collapse unstable core, and thus of the
neutron star newly formed.
Nearly spin-perpendicular kicks of large amplitude 
are required to explain the observations
of geodesic precession in double neutron star binaries such as 
B1913+16. On the contrary,
spin-kick alignment has been claimed for the Vela and Crab pulsars
whose transverse speeds are $\simgreat 70$ and $\sim 170\kms$ respectively.
We suggest that the {\it larger kick component,} when present in a
pulsar,   
results from the formation and disruption
of an evanescent binary, and is   
{\it perpendicular} to the {\it spin axis}; the {\it smaller
kick component} is 
associated some other mechanism that leads to less vigorous kicks, predominantly 
{\it parallel} to the {\it spin} axis because of phase averaging.
This  could give rise to   a 
{\it bimodal distribution in the peculiar velocities
of neutron stars, as it is observed
in the pulsar sample}.
This scenario may explain the {\it run-away black hole}
GRO J1655-40, the first to show evidence for a natal kick.
 
\end{abstract}

\medskip

\section{Introduction}
Radio pulsars have peculiar space velocities
between $\approx 30 \kms$ and $\approx 1600\kms$, significantly greater than
those of their progenitor stars.
The highest speed is from
the Guitar Nebula pulsar (Cordes \& Chernoff 1998),
while the lowest has been recently measured for B2016+28 
by Brisken  et al. (2002)
from very accurate VLBA  pulsar parallaxes.
An early explanation for the
large space velocities of pulsars called for {\it recoil} in a close
binary that becomes unbound at the time of (symmetric) supernova
explosion.  
Now, a number of observations  hint at a {\it natal}
origin of these high space velocities, or at a combination of orbital
disruption and internal kick reaction (when the progenitor
lives in a binary).
Evidence for a kick at the time
of neutron star birth is now found in a variety
of systems, e.g., in double neutron
star binaries such as B1913+16 (Weisberg,
Romani, \& Taylor 1989; Cordes, Wasserman, \& Blaskiewicz 1990; Wex, Kalogera \& Kramer 2000)
where  
misalignment between the spin and orbital 
angular momentum axes indicates velocity asymmetry
in the last supernova.

Among the physical processes that have been proposed to account
for the kick impulses at birth are large-scale 
density asymmetries seeded in the pre-supernova core (leading to
anisotropic shock propagation), asymmetric  neutrino emission
in presence of ultra-strong magnetic fields 
(see Lai, Chernoff, \& Cordes 2001 for
a review), or off-centered electromagnetic dipole emission from the
young pulsar (Harrison \& Tademaru 1975).
However, {\it none of these  mechanisms can explain  kicks 
as large as $\sim 1600 \kms$}.

A further interesting, yet unexplained, statistical property of pulsars
has been pointed out by Arzoumanian, Chernoff and Cordes (2002). They find that 
the pulsar peculiar velocity distribution function is 
{\it bimodal}, i.e., it is described by two almost equally populated 
Gaussian components with
mean velocities of $ 90\kms$ (the {\it fast} pulsars) and $500\kms$ 
(the {\it very fast} pulsars), respectively.
This suggests that two distinct physical mechanisms are
responsible for this bimodality, and in this proceeding we try to
depict a scenario that may account for this dichotomy.

In a recent  paper (Colpi \& Wasserman 2002)
we reconsidered the idea first put forward by Imshennik \&
Popov (1998) that, in the collapse of a rotating core, one or more 
self-gravitating lumps of neutronized matter may form in close orbit
around the central nascent neutron star, transfer mass 
in the short lived binary, and ultimately
explode, 
causing the remaining, massive neutron star to acquire a
substantial kick velocity, as high as the highest observed.
The light member explodes as mass transfer drives it below
the minimum stable mass (against radial perturbations) for a neutron stars.
Stability is lost  upon decompression  
by the $\beta-$decaying  neutrons and nuclear fissions by radio-active
neutron-rich nuclei that deposit energy driving
matter into rapid expansion 
(Colpi, Shapiro \& Teukolsky 1989, 1993; Blinnikov et al. 1984, 1990; 
Sumiyoshi et al. 1998).  
The kick originates from the orbital motion of
this evanescent super-close
binary. Along similar lines Eichler \& Silk (1992) attributed the
largest kicks of pulsars 
to gravitational ejection of fragments formed during rotationally
unstable collapse.

We have studied several effects that may
mitigate the magnitude of the kick, such as gravitational bending
of the exploding debris, and rotational averaging of the momentum impulse.
A point that was previously overlooked is that the minimum mass
of hot still lepton-rich matter is much higher than the value
for cold catalyzed matter, and this opens new evolutionary pathways
(we refer to Colpi
\& Wasserman 2002 for details).

\section {Light fragments around proto-neutron stars}

\subsection {Dynamical rotational non-axisymmetric instabilities}

Formation of a light companion around
a main body implies breaking of spherical and axial symmetry in the
collapse of post-collapse phases. During dynamical collapse, unstable bar
modes can grow in a fluid  that may end with fragmentation. However 
this is known to 
occur only if the core contracts almost isothermally
(Bonnell 1994), but 
collapse in type II supernovae is far from isothermal so that
instabilities of this type do not have time to grow  (Lai 2000).
Can fragmentation/fission be excited after the proto-neutron star core 
has formed ?
{\it Rapid rotation} in equilibrium bodies is known to excite non-axisymmetric
dynamical instabilities. Interestingly, core collapse
simulations of unstable rotating iron cores (Heger, Langer \& Woosley
2000; Fryer \& Heger 2000) or polytropes (Zwerger \& Muller 1997)
indicate that proto-neutron stars, soon after formation, can rotate
differentially above the dynamical stability limit
set when the rotational to gravitational potential energy ratio
$T_{\rm rot}/\vert W\vert$ exceeds  $\sim 0.26.$ 
Strong non-linear
growth of the dominant bar-like deformation ($m=2$) is seen in
these cores (described as polytropes by Rampp,
Muller, \& Ruffert 1998).
The bar evolves, producing two spiral arms that drain the core's
excess angular momentum outwards.
A further possibility that we would like to mention here refers to
the case of {\it fizzlers}, temporary equilibrium bodies supported
almost entirely by rotation, which are 
driven dynamically unstable against bar-like modes
by {\it deleptonization} of hot nuclear matter on timescales of $\approx
1-10$ sec      
(Imamura, \& Durisen 2001), producing spiral patterns around a central core.

In all these above calculations 
however there is no sign of fission/fragmentation into separate condensations. 
How can a double or multiple system form from a 
bar-unstable core or fizzler ?
According to Bonnell's picture (1994), the evolution of the bar instability 
is more complex, in reality.  If the rapidly
spinning proto-neutron star core or fizzler 
goes bar unstable when surrounded by {\it a
fall-back disk}, then matter present in the bar-driven spiral arms 
interacts with this material.  The sweeping of a spiral arm into
fall-back gas can gather sufficient matter to condense, 
by strong cooling and  deleptonization, into a (or a few) fragments of
neutronized matter. This occurs because the $m=1$ mode grows during the
development of the $m=2$ mode. The $m=1$ mode causes the off-centering  
of one spiral arm which then sweeps up more 
material on one side
than the other during {\it continuing accretion}. 
The condensation may eventually
collapse into a low mass proto-neutron star fragment or 
in a few fragments of material. 
This has never been proved numerically.

\subsection {Triggering explosion at the minimum mass limit}

If a light
(proto-)neutron star forms around 
the main central body, 
which limits can be imposed on its mass ? 
Cooling 
plays a key role in addressing these questions.
Goussard, Haensel \& Zdunik (1998), and  Strobel \& Weigel (2001)
have shown that 
the value of the minimum stable mass $m_{\mmc}$ for a neutron star 
(located at the turning 
point in the mass-radius relation of spherical non-rotating equilibria) 
is a function
of the temperature:
it  varies from $\simgreat
1 \msun$ at 50-100 milliseconds after core bounce (setting the actual
value of the mass of the central neutron star), to $\sim 0.7\msun$ after
$\sim 1$ second, down to $\sim 0.3\msun$ after 30 seconds, reaching the
value of $\sim 0.0925\msun$ 
for cold catalyzed matter ($T<1$ MeV)
several hundred seconds later.
Thus, 
the condensation of nuclear matter gathered in the spiral arm by the instability
may become self-bound if its mass $m$ is above the minimum corresponding
to that particular temperature.
Once formed it is stabilized against expansion by cooling.

The value of $m$ and  of the mass ratio $q=m/M$ in the binary 
(with $M$ the heavier of the two 
stars) 
remains  unpredictable to us at this level,  so  
we may just depict three possible scenarios for the formation and evolution 
of this evanescent binary: case ({\bf A}) when the instability sets after
few hundreds 
of milliseconds or a second after core bounce and 
binary formation occurs so that 
$m\simgreat 0.7\msun$ (implying a
pre-existing iron core of  large mass if the primary is as massive as
1.4$\msun$); 
case ({\bf B}) when a lighter star with $m\simeq 0.2-0.4\msun$
forms
around a main body 
after several  seconds or tens of seconds from core bounce or after fizzler 
destabilization (occurring on this time scale); 
case ({\bf C}) when cooling is
sufficiently advanced that the minimum mass approaches its asymptotic
value and the binary companion can have 
$m\simless 0.2\msun$.

\section {Neutron star kicks}

The magnitude of a natal kick in ({\bf A}) is difficult to estimate and 
we must await for realistic simulations of core collapse.  We note
that the evolution in this case might be similar to that described in
coalescing neutron star binaries with  unequal masses (see Rosswog
et al. 2000) where it has been shown that {\it large kicks} can be acquired
as a consequence of  
mass loss via {\it strong winds}.

In cases ({\bf B}) and ({\bf C}), one factor that could lower 
the final neutron star
speed is the finite velocity of the ejecta.  Colpi, Shapiro \& Teukolsky
(1993) have shown that in the dynamical phase of the explosion the
ejecta can attain speeds $w_o$ varying from 10,000 to $\sim 50,000 \kms$. 
These speeds
are close to the escape velocity from the binary and therefore the final
kick imparted to the remaining neutron star may be influenced
substantially by {\it gravitational deflection of the ejecta},
and
{\it velocity phase averaging during the explosion}. A further
complication  is the rapid decay of the orbit separation 
following  {\it unstable mass transfer} and {\it emission of gravitational waves}.
This may lead to coalescence 
before any mass is ejected above escape velocity.

\begin{figure}
\epsfysize=9cm 
\hspace{2.cm} \vspace{0.0cm}
 \epsfbox{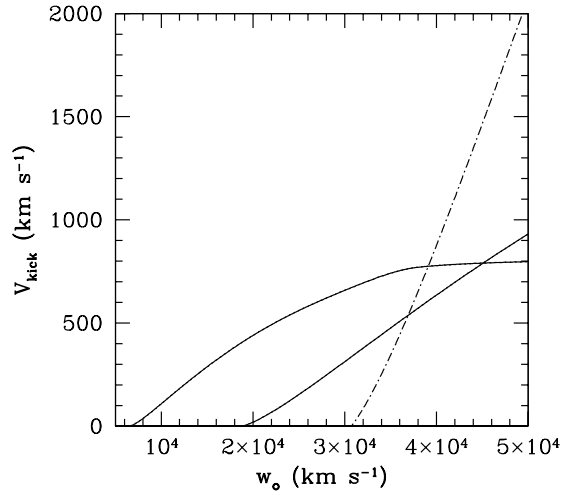}

\caption[h]{Solid lines show the neutron star kick velocity $V_{\rm kick}$ as
a function of the speed of the ejecta $w_o,$ for case ({\bf C}) when the 
light star explodes with a mass 
$m_{\rm expl}\sim 0.3\msun$ (lower curve before crossover) 
and $m_{\rm expl}=m_\mmc=0.09\msun$
(upper curve before crossover). 
The dash-dotted line refers to case ({\bf B}) for a warm star with 
$m_\mmc=0.3\msun.$ 
\label{colpifig1}}
\end{figure}
In {\it binary mass transfer scenarios} (cases {\bf B,C}), 
the {\it maximum kick} velocity 
$V_{\rm kick,max}$ is determined approximately from the value of
the relative orbital velocity between the stars at the 
time the light donor loses its stability $V\simeq ({GM/ r_{\rm
expl}})^{1/2}$ driven by  {\it Roche lobe spill-over}: 
explosion  can occur either 
when the light star reaches $ m_{\mmc}$  after a phase of stable 
mass transfer, 
or at  higher mass  $m_{\rm expl}$ 
if mass transfer is unstable (see Colpi \& Wasserman 2002). 
The maximum kick is thus given by 
\be  
V_{\rm kick,max}\simeq {m_\mmc \over (m_\mmc+M)}\,V\, 
\simeq  {0.68G^{1/2}m^2_\mmc\over 
m_{\rm expl }^{(f-1/6)}[ R(m_{\rm expl})]^{1/2}}M^{-2/3}
\ee
where $R(m_{\rm expl})$ and $m_{\rm expl}$ are the radius and mass of the 
star at the time of explosion ($f<1$.)
{\it Roche lobe spill-over imposes} $V_{\rm kick,max}\propto
M^{-2/3},$ as separation $r_{\rm expl}\propto M^{1/3}$ in contact binaries.  
In case ({\bf B}), the warmer unstable
  proto-neutron star always suffers unstable  mass transfer
and the system may enter a 
phase of common envelope evolution. 
In this case  we expect that the 
star 
develops an unstable core of mass 
$m\approx m_\mmc(T),$ while  losing its envelope, and 
explodes on its own  dynamical time
before  coalescence of the two stars in completed over $\sim$ one  orbital
period.
Ejection of
part of the envelope enshrouding the system 
may provide an additional thrust to the merged object.

In Figure 1, $V_{\rm kick}$ is plotted as a function of the speed
of the ejecta $w_o$ 
for ({\bf B}) and  ({\bf C}), computed including gravitational
bending and orbit phase-averaging. At large $w_o$ the kick approaches its
maximum limiting value. 
As shown in the figure, hydro-dynamical effects produce
{\it a widespread range in velocities}, and most remarkably it is able 
to produce {\it very high kicks,} $V_{\rm kick}\simgreat 1600 \kms$.

\section {On the bimodal distribution of pulsar kick velocities}

The {\it kicks} that result from this mechanism are confined to {\it 
the orbital
plane of the evanescent neutron star binary}, and thus are likely
to be nearly {\it perpendicular} to the {\it
spin 
vector of the remnant neutron star}, which is probably aligned with the
spin of the unstable iron core and the spin of the unstable
post-collapse core.
A spin-perpendicular kick would be
consistent with the requirements imposed by observations of geodetic
precession of B1913+16, where the kick is constrained to lie very
nearly in the plane of the progenitor binary, which was most likely
perpendicular to spins of the spun-up neutron star (i.e. B1913+16)
and its pre-explosion companion star (Wex, Kalogera \& Kramer 2000; 
a kick  $\simgreat 250 \kms$ has been invoked for this system). 

In contrast,  X-ray observations of the Vela pulsar have
revealed a {\it jet parallel to its proper motion} (Pavlov et al. 2000; 
Helfand, Gotthelf \& Halpern 2001), and it has been argued that 
the proper motions of both Vela and the Crab pulsar are closely aligned
with their spin axes. The kick mechanism
studied here would not be able to account for  parallel spin and velocity.
However, we note that the proper motions of both Vela and the Crab correspond
to transverse speeds of $70-141\kms$ and $171\kms$ respectively, using
reasonable estimates of the distances to the pulsars. 
For the alignment to be real, the space
velocities of these two systems must lie in the plane of the sky.
The inferred speeds of these two pulsars are then considerably smaller than
the characteristic speed arising from explosion of a low mass, tidally
disrupted companion. 
Plausibly, the birth of Vela and
the Crab did not involve an evanescent binary phase;
some other mechanism must have been responsible for their spin-aligned
kicks (such as those explored by Lai, Chernoff, \& Cordes 2001).
We are led to suggest that in a pulsar, 
the {\it larger kick component} results from  the formation and disruption of
short-lived super-close binary, and is {\it perpendicular} to the
{\it spin} axis. The {\it smaller kick component} is instead associated with
other {\it less vigorous kicks that tend to align with the rotation axis},
perhaps because of phase averaging (e.g. Spruit \& Phinney 1998). 
A {\it superposition} of these {\it two classes} 
of kicks would also be consistent with the requirement of nearly
but not precisely spin-perpendicular kicks to account for
the observation of geodetic precession in B1913+16.
If this idea is correct, then one expects that {\it lower
velocity neutron stars} should have their {\it space velocities predominantly
along their spin axes}, while {\it higher velocity neutron stars} should have
{\it space velocities predominantly perpendicular to their spins}. This would 
create  {\it two nearly 
independent components} in the velocity distribution: the ``fast'' and ``very
fast''
neutron stars.
A contamination of low velocity stars belonging to the low velocity
``tail''
of the high velocity component would come from those explosions
in the evanescent binary where light bending and rotational averaging
have been more important.
This could imply the occurrence  of {\it bimodal} distribution of velocities 
for the pulsar population as such 
inferred from the observations (Arzoumanian, Chernoff, \& Cordes 2002).
Since the two classes of ``fast'' and ``very fast'' pulsars
are nearly equally populated, pulsars with large kicks are not exceptional.
This would imply that progenitor stars with rapidly rotating iron cores 
are common in Nature.

\section {Black hole kicks}
Recently it has been reported that the binary X-ray 
Nova GRO J1655-40 (Mirabel et al. 2002) hosts the first black hole candidate 
for which we have evidence of a {\it runaway motion} of $112\pm 18 \kms$ 
imparted by a {\it natal} kick. GRO J1655-40 is a source 
that shows $\alpha$-elements in its optical spectrum indicating that formation
of the black hole was accompanied by a supernova explosion (Isrealian et
al. 1999). The black hole was very likely born in a two step process in which
a hot proto-neutron star forms first, accompanied by the launch of a successful
shock wave, but is subsequently destabilized to gravitational collapse by 
substantial fallback of material from the stellar envelope.

The evanescent binary scenario, proposed here, for the origin of kicks
applies naturally to
the black hole case, particularly in a fall-back scenario (and not
for prompt core collapse).
Continuing accretion onto the off-centered spiral pattern could be even more 
compelling in this case, leading to the formation of a massive fragment, 
and substantial fall-back on the side where the pattern is not develop fully.
Thus, if the formation
of a light exploding proto-neutron star accompanies core
collapse to a black hole, its maximum kick velocity would scale as
$V_{\rm kick, bh}=
V_{\rm kick, ns}(M_{\rm ns}/M_{\rm bh})^{2/3}~.$ 
If we hypothesize that both fast black holes and ``very fast'' pulsars
acquire their large speeds via the disruption of an evanescent binary,
then we infer fast black hole recoil velocities
\begin{equation}
V_{\rm kick, bh}\sim 170 \kms \left ({
V_{\rm kick, ns}\over 500 \kms}\right )\left ({M_{\rm ns}\over 1.3\msun}
{5.4 \msun \over M_{\rm bh, GRO J1655}}\right )^{2/3}~.
\end{equation}
We note that a kick of this magnitude is consistent with the observed value
for GRO J1655-40 (Mirabel et al. 2002).

\end{document}